\pdfoutput=1

\documentclass[sigconf, authorversion]{acmart}

\usepackage{multicol}
\usepackage{enumitem}
\usepackage{todonotes}
\usepackage{multirow}
\usepackage{natbib}
\usepackage{wasysym}
\usepackage{arydshln}
\usepackage{soul}

\usepackage{graphicx}
\usepackage{caption}
\usepackage{subcaption}

\AtBeginDocument{%
  \providecommand\BibTeX{{%
    \normalfont B\kern-0.5em{\scshape i\kern-0.25em b}\kern-0.8em\TeX}}}

\makeatletter
\newcommand\footnoteref[1]{\protected@xdef\@thefnmark{\ref{#1}}\@footnotemark}
\makeatother

\usepackage{caption}
\usepackage{subcaption}

\copyrightyear{2025}
\acmYear{2025}
\setcopyright{rightsretained}
\acmConference[Australasian Computing Education Conference]{Submitted to ACE 2025}{February 2025}{Brisbane, Australia}
\acmBooktitle{ACE '25}
\acmDOI{}

\begin{document}




\title[Breaking the Programming Language Barrier]{Breaking the Programming Language Barrier: Multilingual Prompting to Empower Non-Native English Learners}






\author{James Prather}
\orcid{0000-0003-2807-6042}
\affiliation{
  \institution{Abilene Christian University}
  \city{Abilene, Texas}
  \country{USA}
}
\email{james.prather@acu.edu}

\author{Brent N. Reeves}
\email{brent.reeves@acu.edu}
\orcid{0000-0001-5781-1136}
\affiliation{%
  \institution{Abilene Christian University}
  \city{Abilene, Texas}
  \country{USA}
}

\author{Paul Denny}
\orcid{0000-0002-5150-9806}
\affiliation{
  \institution{The University of Auckland}
  \city{Auckland}
  \country{New Zealand}
}
\email{paul@cs.auckland.ac.nz}

\author{Juho Leinonen}
\orcid{0000-0001-6829-9449}
\affiliation{
  \institution{Aalto University}
  \city{Espoo}
  \country{Finland}
}
\email{juho.2.leinonen@aalto.fi}

\author{Stephen	MacNeil}
\orcid{0000-0003-2781-6619}
\affiliation{
  \institution{Temple University}
  \city{Philadelphia}
  \state{PA}
  \country{United States}}
\email{stephen.macneil@temple.edu}

\author{Andrew Luxton-Reilly}
\orcid{0000-0001-8269-2909}
\affiliation{
  \institution{The University of Auckland}
  \city{Auckland}
  \country{New Zealand}
}
\email{a.luxton-reilly@auckland.ac.nz}

\author{Jo\~{a}o Orvalho}
\orcid{0000-0002-9185-4479}
\affiliation{
  \institution{Polytechnic University of Coimbra}
  \city{Coimbra}
  \country{Portugal}
}
\email{orvalho@esec.pt}

\author{Amin Alipour}
\orcid{0000-0002-6479-7202}
\affiliation{
  \institution{University of Houston}
  \city{Houston}
  \state{Texas}
  \country{United States}
}
\email{maalipou@central.uh.edu}

\author{Ali Alfageeh}
\orcid{0009-0001-9106-0433}
\affiliation{
  \institution{University of Houston}
  \city{Houston}
  \state{Texas}
  \country{United States}
}
\email{aalfagee@cougarnet.uh.edu}

\author{Thezyrie Amarouche}
\orcid{0000-0003-3725-0049}
\affiliation{%
  \institution{University of Toronto Scarborough}
  \city{Toronto}
  \country{Canada}
}
\email{thezyrie.amarouche@mail.utoronto.ca}

\author{Bailey Kimmel}
\orcid{0009-0000-6655-0564}
\affiliation{
  \institution{Abilene Christian University}
  \city{Abilene, Texas}
  \country{United States}
}
\email{blk20c@acu.edu}

\author{Jared Wright}
\orcid{0009-0000-0255-8989}
\affiliation{
  \institution{Abilene Christian University}
  \city{Abilene, Texas}
  \country{United States}
}
\email{jpw19b@acu.edu}

\author{Musa Blake}
\orcid{0009-0001-2457-9147}
\affiliation{
  \institution{Abilene Christian University}
  \city{Abilene, Texas}
  \country{United States}
}
\email{mbb23c@acu.edu}

\author{Gweneth Barbre}
\orcid{0009-0007-3125-4168}
\affiliation{
  \institution{Abilene Christian University}
  \city{Abilene, Texas}
  \country{United States}
}
\email{gab23c@acu.edu}

\renewcommand{\shortauthors}{Prather et al.}

\begin{abstract}

Non-native English speakers (NNES) face multiple barriers to learning programming. These barriers can be obvious, such as the fact that programming language syntax and instruction are often in English, or more subtle, such as being afraid to ask for help in a classroom full of native English speakers. However, these barriers are frustrating because many NNES students know more about programming than they can articulate in English. Advances in generative AI (GenAI) have the potential to break down these barriers because state of the art models can support interactions in multiple languages. Moreover, recent work has shown that GenAI can be highly accurate at code generation and explanation. In this paper, we provide the first exploration of NNES students prompting in their native languages (Arabic, Chinese, and Portuguese) to generate code to solve programming problems. Our results show that students are able to successfully use their native language to solve programming problems, but not without some difficulty specifying programming terminology and concepts. We discuss the challenges they faced, the implications for practice in the short term, and how this might transform computing education globally in the long term.


\end{abstract}




\maketitle

%
%
\section{Introduction}
\label{sec:intro}

Non-native English speakers (NNES) face significant challenges when learning programming through English language instruction. They tend to set higher academic goals for themselves and spend more time studying compared to their native English-speaking peers \cite{guzman2021experiences}. The stance that one must know English to be a programmer is quite entrenched in the broader community\footnote{Reinforced by documents such as PEP 8 -- a style guide for Python -- with guidance such as~\textit{Python coders from non-English speaking countries: please write your comments in English}.}, but this view is often critiqued \cite{hanselman2008doyou,mcculloch2019coding}. The impacts of learning English-adjacent programming languages through English language instruction are not well understood. Although NNES sense of belonging does not appear to be impacted by learning programming in English, other underlying factors can lead to a less inclusive environment. On the one hand, these students often express higher self-doubt and embarrassment by asking for help \cite{agarwal2022analysis, molina2024leveraging}. 
On the other hand, students enjoy learning programming in their native language and report that it positively impacts their experience \cite{raj2017whatdo}. However, instruction in native language does not seem to impact learning outcomes for NNES students \cite{raj2018doesnative,agarwal2022analysis}. While one might think that English language instruction would vary across the globe being sensitive to local contexts, a recent study showed that it was remarkably monolithic across three different continents \cite{becker2022horses}. 

In 2019, Becker wrote that native natural language programming would have obvious advantages for NNES but remained a far-off fantasy \cite{becker2019parlezvous}. The advent of generative AI (GenAI) since then has radically shifted our perception of the role that natural language might play in programming~\cite{denny2024computingCACM, prather2023robots}. Initial work on generating programming exercises using GenAI found that it could create coherent and customizable programming assignments \cite{sarsa2022automatic}. Follow-up work showed that students in Finland preferred customized programming assignments, albeit in English \cite{logacheva2024evaluating}. Recent work has demonstrated the capabilities of ChatGPT-3.5 to generate programming problems not only in English but also in Tamil, Spanish, and Vietnamese \cite{jordan2024need}. Other types of programming problems,  such as the classic ``Explain in Plain English'' (EiPE), are also being automated by GenAI that can be used to grade them at scale \cite{denny2024explaining, kerslake2024integrating}. 
These advancements offer a unique opportunity to explore expressing programming concepts in a learner's native language.

In this paper, we present the first exploration of NNES prompting GenAI to solve programming problems using their native language. We do this by utilizing Prompt Problems with prompts in a learner's native language. Prompt Problems are a type of programming exercise designed to teach programming concepts via GenAI prompting. Students receive a problem visually and must write a prompt that can generate code to solve the problem. If the generated code does not pass the suite of test cases, then they must edit their prompt again, iterating until it successfully passes. Initial work has shown that students find Prompt Problems engaging and encourage metacognitive reflection \cite{denny2024promptproblems}. Here, we combine several threads of recent work to see if students can solve these problems in their native languages (Chinese, Portuguese, Arabic) and how it impacts their learning experience.

Therefore, our research questions are:
\begin{enumerate}
    \item \textbf{RQ1:} How successful are students at solving Prompt Problems in their non-English native languages?
    \item \textbf{RQ2:} How does solving Prompt Problems in non-English native language impact user experience?
\end{enumerate}

%
%
\section{Related Work}
\label{sec:related-work}

\subsection{LLMs in Computing Education}

Since the popularization of large language models (LLMs) with the release of ChatGPT in November 2022, researchers in computing education have explored both the opportunities and challenges presented by these models. 


\subsubsection{Opportunities}

One of the earliest identified opportunities of LLMs for computing education is the automatic generation of programming exercises. Sarsa et al. studied the capabilities of Codex in generating novel programming exercises focusing on specific themes (such as `cooking' or `basketball') and concepts (such as `for-loops')~\cite{sarsa2022automatic}. Their findings suggested that LLMs can create novel exercises that are often good enough to provide to students as-is. Follow-up work has found that while LLMs are good at generating high-quality exercises~\cite{logacheva2024evaluating,gutierrez2024evaluating}, the thematic contextualization of created content is often shallow~\cite{logacheva2024evaluating}.

In addition to programming exercises, LLMs have been used to create multiple-choice questions (MCQs) for computing courses~\cite{agarwal2024understanding,tran2023generating,doughty2024comparative}. Tran et al. found promising results when using GPT-4 to generate multiple-choice questions that were isomorphic to provided examples ~\cite{tran2023generating}.
In similar work, Doughty et al. examined GPT-4's performance for generating MCQs that would align with course learning objectives~\cite{doughty2024comparative}. Their results suggest that GPT-4 was able to generate very high-quality MCQs that were evaluated to be of similar quality as those generated by human educators.

Researchers have also explored using LLMs to help students understand code. Recent work by Bernstein et al. had students create personally meaningful analogies for recursion using ChatGPT, a notoriously difficult threshold concept~\cite{bernstein2024like}. The analogies created by students with the help of ChatGPT were more diverse compared to generic analogies that LLMs would generate on their own. Additionally, the participants reported that these analogies helped them understand recursion.

LLMs have also been used to explain both program code~\cite{sarsa2022automatic,macneil2023experiences,leinonen2023comparing,jury2024evaluating} and programming error messages~\cite{leinonen2023using,wang2024large,taylor2024dcc,santos2024not}. Sarsa et al. found that Codex was able to explain program code in natural language correctly for approximately two thirds of the lines of code in the examples they provided it~\cite{sarsa2022automatic}. MacNeil et al. reported that students found LLM-generated code explanations useful and helpful for their learning~\cite{macneil2023experiences}. Leinonen et al. found that code explanations generated by LLMs were rated as being more accurate and easier to understand compared to student-generated explanations~\cite{leinonen2023comparing}. For programming error messages, while Codex was only sometimes helpful in enhancing them~\cite{leinonen2023using}, follow-up work using GPT-3.5 and GPT-4 has found improved results. Taylor et al. found that GPT-3.5 explained errors correctly in up to 90\% of cases~\cite{taylor2024dcc}, and Wang et al. found that students who were provided enhanced programming error messages created by GPT-4 repeated errors less frequently and resolved errors with fewer attempts~\cite{wang2024large}. However, Santos et al. found that the time to fix errors was not improved by providing GPT-4 enhanced error messages~\cite{santos2024not}.

LLMs have also been explored for providing direct assistance to students, either through responses to help requests~\cite{hellas2023exploring} or by feedback on program code~\cite{koutcheme2024open,kiesler2023exploring}. Hellas et al. explored how well LLMs could respond to students' help requests and found that while GPT-3.5 would often detect issues in student's incorrect code, it would also hallucinate nonexistent issues~\cite{hellas2023exploring}. Kiesler et al. found that ChatGPT was able to detect and correct compiler errors well, but had lower performance for other types of errors ~\cite{kiesler2023exploring}. Koutcheme et al. used multiple smaller, open-source LLMs and found that some of the open-source models such as Zephyr-7B-beta rivaled some proprietary models such as GPT-3.5 in performance ~\cite{koutcheme2024open}.


\subsubsection{Challenges}

Despite the many opportunities that LLMs provide, there are challenges too. The earliest computing education work that utilized LLMs found that they can solve most introductory programming (CS1) exercises~\cite{finnieansley2022robots}, and follow-up work with more recent models has demonstrated rapidly improving performance ~\cite{prather2023robots}. 
Their problem solving is not limited to programming exercises, as LLMs have been found to solve Parsons problems too~\cite{reeves2023evaluating}, even if provided as images~\cite{hou2024morerobots}, as well as computer graphics questions requiring both visual perception and geometric reasoning skills \cite{feng2024eye}. LLMs can also solve multiple-choice questions related to programming~\cite{savelka2023thrilled}. This has raised fears that students might use LLM support for academic misconduct, or over-rely on LLM support even when their intention is not to cheat. While some work has looked into automatically detecting LLM-generated solutions~\cite{orenstrakh2024detecting,hoq2024detecting}, there are no sure-fire methods to conclusively classify code as LLM-generated.

Perhaps more concerning is that students could inadvertently be negatively affected by LLM use. Prather et al. replicated an earlier study~\cite{prather2018metacognitive} looking into metacognitive difficulties that novice programmers face while they're programming, but this time giving students access to LLM tools such as GitHub Copilot and ChatGPT~\cite{prather2024widening}. They found that not only did students still face the same metacognitive difficulties, but some of these were exacerbated by the LLMs and new difficulties were also introduced. While the best students were able to accelerate with the help of GenAI tools, students who were already struggling faced more difficulties, which could widen the gap between the best and the worst performing students. This result is similar to help-seeking, where Hou et al. found that students who are capable at using models get the most benefits~\cite{hou2024effects}.
These issues highlight the need for novel pedagogies that help students to learn to successfully leverage GenAI models in their work.






\subsection{Emerging LLM-Powered Pedagogies}

Teaching students to write natural language prompts is an emerging area in computing education, and can be applied equally well to support both code writing and code comprehension tasks. Moreover, given the language translation capabilities of LLMs, pedagogies based on this idea could be applicable in any spoken language, potentially improving the accessibility of programming education for diverse student populations.  

\subsubsection{Prompt Problems}

When exploring the performance of LLMs for solving typical CS1 problems, Denny et al. observed that making certain refinements to the prompts led to greater accuracy in the generated code \cite{denny2023conversing}.  They argued that learning how to craft effective prompts is essential for novice programmers, but did not directly propose teaching strategies for developing this skill.  In subsequent work, Denny et al. introduced the idea of `Prompt Problems' as a novel exercise for helping students learn how to solve computational tasks through natural language prompts \cite{denny2024promptproblems}.  In a Prompt Problem, a student would be shown a computational task, typically presented visually without any textual description, and they would write a natural language prompt for an LLM to generate code to solve that task.  Prompt Problems allow shifting the focus from syntax mastery, often emphasized early in programming education, towards higher-level problem-solving.  The authors presented and evaluated a tool called Promptly which would execute the generated code against predefined test cases to assess both the effectiveness of the prompt and the student's understanding of the problem.  Feedback on the activity from students in both a CS1 and CS2 course suggested that they appreciated that it engaged their computational thinking skills and introduced them to new programming constructs.  However, it also highlighted the need for further research into how these exercises could be best integrated into classroom practice and whether they could meaningfully improve learning outcomes.  In addition, all prior research involving Prompt Problems has been in the context of English-language prompting.

\subsubsection{Explain in Plain Language (EiPL)}

While Prompt Problems can be viewed as an alternative approach to solving \emph{code writing} tasks, by evaluating code generated from a student's prompt against a test suite, an analogous approach can be used to provide feedback on \emph{code comprehension} tasks.  There is a considerable body of research in the computing education literature on `Explain in Plain English' (EiPE) questions, which require students to articulate the purpose of a code fragment in natural language.  Early work exploring this type of question by Whalley et al. identified a connection between students' programming skills and their ability to describe code accurately in simple terms \cite{whalley2006australasian}.  Over the following decade, further research validated this approach across various contexts \cite{lopez2008relationships, venables2009closer, lister2009further}.  In 2014, Corney et al. highlighted that EiPE tasks enhance students' ability to reason about code, which can, in turn, improve their coding skills \cite{corney2014explain}, however scaling these exercises has been challenging due to the subjective nature of grading free-text responses. 


Applying LLMs to the task of grading EiPE responses has shown great promise \cite{denny2024explaining}.  Smith et al. proposed `Code Generation Based Grading' (CGBG) \cite{smith2024cgbg}, a method conceptually similar to the idea for generating feedback on Prompt Problems, by using LLMs to generate code based on a student's EiPE response.  The generated code is then tested against predefined test cases, offering both objective grading and actionable feedback for students.  Their study demonstrated that CGBG aligns well with human grading while providing a scalable solution for large classes, making it a promising approach for automated assessments.  Smith et al. further explored the connection between prompt writing and code comprehension \cite{smith2024prompting}. Their research showed that students benefit from the dual challenge of writing prompts and understanding generated code. By performing both tasks, students not only develop their code comprehension skills but also gain proficiency in prompt engineering. Similar recent work has shown that LLM-grading of EiPE questions is engaging for students, but importantly also found that relational responses, 
where students integrated code elements into high-level summaries, were the most successful in generating correct code \cite{denny2024explaining}.  This line of research highlights the potential for LLM-powered exercises to foster higher-order thinking, and to help students appreciate the connection between natural language prompts and the code they generate.   Very recently, Kerslake et al. examined the integration of both Prompt Problems and EiPE questions into a large introductory programming course \cite{kerslake2024integrating}. A key finding of this work was that students who struggled with traditional coding tasks performed better with natural language prompts, suggesting that these tasks engage a broader range of cognitive skills and provide a more accessible entry point for novice programmers.

\subsubsection{Multilingual Support in Programming Education}

Non-native English-speaking students often face significant challenges in programming education \cite{agarwal2022analysis, guo2018nonnative, guzman2021experiences}. For example, Becker highlighted how the predominance of English in programming languages, documentation, and error messages adds a significant cognitive load to non-native speakers, making it harder for them to fully participate in programming courses \cite{becker2019parlezvous}. This study revealed that non-native speakers often encounter difficulties interpreting keywords and comments, increasing their cognitive load and limiting their ability to focus on problem-solving.  Although instructors have been shown to adjust their speech patterns and vocabulary to meet the needs of diverse student groups \cite{becker2022horses}, modern LLMs have suddenly opened the door to providing significant multilingual support in computing education.  Recent work by Jordan et al. explored the use of LLMs to generate programming exercises in Spanish, Vietnamese, and Tamil \cite{jordan2024need}.  Although the model they used at the time of their research is no longer state of the art, support for generating resources in Spanish and Vietnamese was generally good.  Despite some limitations, which will likely lessen over time as model capabilities improve, the authors see great potential in using LLMs as a pathway for creating culturally relevant programming resources tailored to non-native speakers.  This is a particularly important result given that when students see their identities reflected in learning activities, they experience positive academic and social outcomes, making computer science more relevant and accessible to them \cite{Jacob2020}.

There has recently been some early exploration of multilingual support for prompting-based activities in computing education.  Smith et al. explored the use of EiPE questions in Indic languages such as Hindi, Tamil, and Marathi, revealing both opportunities and challenges in supporting multilingual learners \cite{smith2024explainplainlanguagequestions}. While students appreciated the ability to engage with programming tasks in their native languages, many still preferred English for technical precision and familiarity. This suggests that while multilingual support can lower accessibility barriers, students' existing preferences and technical contexts must also be considered when designing programming exercises.

The relationship between programming and natural languages has been further explored by Veldthuis and Hermans who applied natural language vocabulary acquisition models to programming education \cite{veldthuis2024word}. Their study restructured introductory programming lessons to include strategies typically used in second language learning, such as scaffolding vocabulary -- the gradual introduction of new concepts in phases, with opportunities for practice and verification. Using the Hedy programming environment, students were progressively exposed to increasing complexity, initially permitting errors in syntax to mirror natural language learning processes. The researchers found that these strategies not only enhanced students' understanding of programming concepts but also improved engagement and motivation. This approach aligns with the broader goal of our present study, which investigates how LLMs can support multilingual Prompt Problems by offering adaptive, language-aware programming exercises that accommodate students’ linguistic backgrounds and lower cognitive barriers to learning.

%
%
\section{Methodology}
\label{sec:methods}
We explored the use of Prompt Problems in three distinct environments: a university with English as the language of instruction, but where NNES students completed tasks using Chinese; a Portugese university where students completed tasks using Portugese; and, a Middle-Eastern university in which students used Arabic.  In each institution, the prompts used by students were collected along with a post-survey.

\begin{table*}[ht]
\caption{Prompt Problems used in the three studies}
\label{tab:promptproblems}
\centering
\begin{tabular}{|p{0.15\linewidth}|p{0.45\linewidth}|p{0.25\linewidth}|}
\hline
\textbf{Name} & \textbf{Description} & \textbf{Example} \\ \hline
\texttt{1. scramble} & Write a Python function called \texttt{scramble} that accepts a string containing only lowercase letters and a number as a parameter. It returns a string with each letter shifted by the number. & \texttt{scramble("zoo", 2) => 'bqq'} \\ \hline
\texttt{2. arrange} & Write a function in Python called \texttt{arrange} that accepts a string and returns a new string with the letters sorted: capital letters in ascending order followed by lowercase letters in descending order. & \texttt{arrange("AaBbCcDd") => 'ABCDdcba'} \\ \hline
\texttt{3. speak} & Write a Python function called \texttt{speak} that replaces characters with numbers: 'e' -> '3', 'o' -> '0', 's' -> '5', 't' -> '7', 'a' -> '4', 'i' -> '1', including their uppercase equivalents. & \texttt{speak("Hello World!") => 'H3ll0 W0rld!'} \\ \hline
\end{tabular}
\end{table*}


\subsection{Data Collection}
The first data collection was conducted in February 2024 at Polytechnic University of Coimbra in Portugal with 27 students whose native language was Portuguese. Students were in a CS2 course learning Python and instructed in Portuguese.
The second data collection was conducted in May 2024 at the University of Auckland, New Zealand, in a post-graduate course with 19 students using Python. The language of instruction was English, but
most of the students enrolled in the course were non-native English speakers whose primary language was Chinese. 
The third data collection was also in May 2024 at Umm Al-Qura University in Saudi Arabia with 34 students. These students were undergraduates learning in Java and the language of instruction was Arabic.
In each context, the instructor explained the concept of Prompt Problems beforehand and it was demonstrated to them. Students were then invited to participate by completing the Prompt Problems shown in Table~\ref{tab:promptproblems} by prompting in their native (i.e. non-English) language. After the activity, a short post-survey was administered to capture their experience solving Prompt Problems in their native language.

\subsection{Analysis}
We performed a quantitative analysis of the prompt submission data (RQ1) and a qualitative analysis of the survey data (RQ2).

\subsubsection{Quantitative Analysis}
The initial dataset included 1,771 total prompts by students in all three groups. Because we were only interested in the prompts by NNES students, we cleaned the data by removing all prompts by students who only prompted in English. While it's possible that a NNES student decided to only prompt in English, our research questions are directed at the experience of students prompting in their native language. Therefore, all of the 1679 remaining prompts contained at least some non-English words. We then calculated completion data for each problem for each language group.

After visually examining the text of the remaining prompts, two researchers created six categories of prompts based on the amount of English, native language, syntax, or code in the prompt. Four other researchers then categorized every prompt into one of those categories. If there was concern about which category to place a prompt into, the four researchers categorizing discussed the prompt and came to consensus. These decisions were randomly sampled for quality assurance purposes by the first two researchers who made the categories. The groups were as follows:

\begin{itemize}
  \item \textbf{O:} Other (empty prompt, URL, etc.) 
  \item \textbf{E:} English, except for a single non-English word or very short phrase 
  \item \textbf{N:} Non-English language except for computer language terms / jargon (examples include: string, char, array, function, def, function names, result, for, repeat, in, text, change, package, import, etc.) 
  \item \textbf{M:} Mixed
  \item \textbf{C:} All code
\end{itemize}

These categories represent the kinds of linguistic prompting strategies that students used to solve the Prompt Problems. We analyzed these data to find the most common strategies used by different language groups as well as the kinds of state changes that students went through while solving the problem.

\subsubsection{Qualitative Analysis}


The data was coded by a member of the research team following a reflexive thematic analysis as outlined by Braun and Clarke~\cite{braun2019reflexive}. This approach highlights the researcher's active role in the development of themes. This inductive approach is particularly suited to exploratory research with smaller, diverse samples. Given the relatively small sample size and the heterogeneity between the three participant groups, an inductive strategy facilitated the generation of themes grounded in the data, rather than being constrained by preconceived theoretical frameworks.

The primary goal of the analysis was therefore to identify patterns and trends that can contextualize our quantitative findings and to inform new hypotheses. Unlike in positivist methods, where sample size is predetermined based on statistical power, thematic analysis does not require a set number of participants. Instead, our process was guided by thematic saturation, which is a point at which new responses cease to produce new insights~\cite{fusch2015we}. In our analysis, thematic saturation was considered with regards to each participant linguistic group. Each new sample provided additional insights, but within groups, saturation was often observed within the first 15 responses. Responses consistently aligned with the primary themes that emerged early in the coding process for each sample, and despite minor variations in individual responses, no new dominant themes were identified during the latter stages of analysis. A limitation that arises from this analysis is that our findings are suggestive, but not conclusive or capable of generalizing beyond the participants sampled in this study.


\begin{table}
\small
\caption{Summary of student success rate by language and question}
\begin{tabular}{c|ccc}
\label{tab:quant-completion-summary-2}
\textbf{Language} & \textbf{Problem 1} & \textbf{Problem 2} & \textbf{Problem 3} \\
 \phantom{x} & \multicolumn{3}{c}{Students Attempting / Correct } \\
\toprule
Arabic & 42 / 27 & 26 / 15 & 14 / 12 \\
Chinese & 12 / 12 & 12 / 5 & 4 / 2 \\
Portuguese & 22 / 21 & 21 / 20 & 19 / 18 \\

\bottomrule
\end{tabular}
\end{table}

%
%
\section{Results}
\label{sec:results}

\subsection{Quantitative Data}
The data presented in Table \ref{tab:quant-completion-summary-2} show that a majority of students were able to complete the exercises using at least some of their native language. As noted in previous work on Prompt Problems, the sequential nature of the tasks means there are naturally fewer students who attempt subsequent problems \cite{denny2024promptproblems}. 

\begin{table}
\small
\caption{The total number of prompts that used a particular strategy with the number of correct prompts that used that strategy in parenthesis.}
\begin{tabular}{c|ccccc}
\label{tab:quant-category-summary}
\textbf{Language} & \multicolumn{5}{c}{\textbf{Linguistic Prompting Strategy Counts}} \\
\phantom{x} & Other & English & Native & Mixed & Code \\
\toprule
Arabic     & 4 (0) &  7 (0) & 1123 (55) &  18 (2) & 0 (0) \\
Chinese    & 0 (0) & 26 (6) &    40 (6) &  22 (5) & 4 (2) \\
Portuguese & 0 (0) &  4 (2) &  428 (76) &   2 (2) & 0 (0) \\ 
\bottomrule
\end{tabular}
\end{table}

The data from the categorization (see Table \ref{tab:quant-category-summary}) show that most students attempted to prompt in their native language (category ``N''). The next highest categories were ``M'' (Mixed language) and ``E'' (almost all English except for one native language word or phrase). Given that students were able to solve the Prompt Problems (see Table \ref{tab:quant-completion-summary-2}) and that most prompted in their native language almost exclusively, these data show that students were able to successfully solve the Prompt Problems in their native language. However, there is a clear divide between correct solutions in Chinese and Portuguese compared to Arabic. For category ``N'', Chinese had a 15\% success rate with that prompting strategy and Portuguese had an 18\% success rate. Prompts that were in Arabic, however, had a 5\% success rate. Interestingly, students prompting in Chinese resorted to using English the most and found some success with it.


The vast majority of submission streams (prompt submissions attempting to answer one of the Prompt Problems) used one category exclusively. We had expected to find many more submission streams that included more than one state. Of 172 successful submission streams, 152 used only 1 state, and 127 of those were \textit{Native}. Nine streams used two states, meaning that, for example, after submitting a prompt that was categorized as \textit{Native}, the student switched to, for example, \textit{English}.  Of those streams that had two states, most were \textit{English}-then-\textit{Native} or \textit{Native}-then-\textit{English}. Nine used three, one used five, and one used six states.  \textit{Native} was the most frequent starting state as well as the most frequent ending state.  





\subsection{Qualitative Data}

Across the survey responses, three primary themes were identified regarding the use of native versus foreign language. First, participants shared that models generally performed poorly when interacting in their native language. This poor performance appeared to be more pronounced for the Arabic-speaking sample, a finding which can be seen in the quantitative data above. Second, participants highlighted trade-offs between using their native language and communicating in English, often finding English leads to better results, despite being less expressive for them. Finally, some participants mentioned that while their native spoken language was not English, they were more accustomed to coding in English, suggesting it felt more natural to use English in the context of programming. In the following subsections, participants are labeled by language: PA = Participant Arabic, PC = Participant Chinese, PP = Participant Portuguese.

\subsubsection{Poor Support for Some Languages} 


A recurring theme across all participant samples was the inadequate support LLMs provide for some languages compared to others.  This observation is consistent  with prior research in natural language processing, which refers to these languages as `low-resource'---languages that have less data available in the original model's training set \cite{magueresse2020low}. For example, PA-1 explained how ChatGPT can face challenges when understanding the Arabic language: 

\begin{quote}
    \textit{``There are challenges with ChatGPT's understanding of the Arabic language.''} (PA-1)
\end{quote}

Anecdotally, this theme was much more prevalent in responses in the Arabic-speaking sample than in the Chinese-speaking sample. This again relates to the concept of `low-resource' languages as 
there is currently more Chinese text included in training sets for large language models. 


One participant speculated that this might stem from code often being written in English, which creates a gap when using other languages to interact with code or to explain technical concepts. 

\begin{quote}
    \textit{``In English, it's easy to write what you want because the programming is in English, while in Arabic, it was a bit difficult to convey the information to ChatGPT''} (PA-4) 
\end{quote}

Due to this poor performance for native languages, multiple participants expressed an explicit preference for using English rather than their native language when interacting with language models.

\begin{quote}
    \textit{``It was somewhat bad, but I think it would be much better if it were in English.''} (PA-14)
\end{quote}

Interestingly, some participants implied that using their second language, English, had an unexpected benefit. They described how writing prompts in English required them to be more deliberate and considerate about the language they used, which forced them to slow down and organize their thoughts more carefully. 

\begin{quote}
    \textit{``I'm so familiar with native language, so the advantage is I can write the command more smoothly, while this makes my command omit some points, which makes the model hard to understand. [To] use English, I need to organize my language and always consider about the grammar, so it might be a little bit slow, but accurate.''} (PC-16)
\end{quote}

This observation suggests that slowing down and planning may offer metacognitive benefits, as planning is an important metacognitive strategy. However, more research is needed to systematically investigate this potential effect.

Despite the poor performance claimed by many participants, a small minority of participants expressed a preference for their native language. For example PA-24 explained that Arabic was easier but required more effort for them to explain the requirements to the model: 

\begin{quote}
    \textit{``The Arabic language is easier, but it requires a lot of explanation.''} (PA-24)
\end{quote}

\subsubsection{Trade-offs between Expressivity and Model Performance}

Participants across the samples frequently shared about the trade-offs they face being more expressive in their native language and achieving  better model performance when using English. Many participants used the word `expressive' to describe how much easier it was to communicate their intent and goals in their native language. For example, PC-12 explained:

\begin{quote}
    \textit{``Native language is more easy for me to express my thoughts. But more difficult for ChatGPT to understand me.''} (PC-12)
\end{quote}

And this experience was also shared by PA-4 in the Arabic-speaking sample: 

\begin{quote}
    \textit{``The advantages are that Arabic is my language, so I will have strong expression in it. However, the disadvantage I faced is that the program did not understand the explanation well.''} (PA-4)
\end{quote}

This trade-off was not unique to one language. For instance, PA-33 and others shared similar experiences about how language models did not appear to understand some of the words from their native language: 

\begin{quote}
    \textit{``Since it is my native language, it was much easier than English. [However,] it didn't understand some Arabic words.''} (PA-33)
\end{quote}

Another participant noted the balance between providing more detailed descriptions in their native language and achieving greater accuracy in English:

\begin{quote}
    \textit{``I can give a more detail description with my native language, but seems english description can be more accurate''} (PA-2) 
\end{quote}

PC-15 attributed some of this trade-off to differences in meaning between the two languages: 

\begin{quote}
    \textit{``When I use my native language, I can express my ideas more accurately and avoid grammar problems to some extent. However, some of the same words may have very different meanings in my native language and in English, and also, many of the inflectional structures are completely different, which causes some problems.''} (PC-15)
\end{quote}

One participant speculated that the complexity of their descriptions in their native languages might have contributed to the model misunderstanding them. For example, PA-1 said:  

\begin{quote}
    \textit{``The abundance of vocabulary made it difficult for it to understand''} (PA-1)
\end{quote}

There appears to be a trade-off between the ease and clarity of expressing one's thoughts in their native language and the performance benefits gained by code-switching to English. While participants found it more natural to articulate ideas in their own language, they believed using English provided better comprehension and accuracy. 
However, this trend was not seen among responses by Portuguese students. Nine students used the word ``easy'' and 13 used the word ``positive'' in their responses. One example is PP-15:

\begin{quote}
    \textit{``Solving the exercises with Promptly was easy, I didn't have any difficulties and I found it quite intuitive.''} (PP-15)
\end{quote}

This may be because some languages, like European ones, would be considered ``high-resource'' languages and would therefore not suffer the same expressivity trade-offs as languages with non-Latin alphabets. The quantitative data correlates to this trend in the qualitative data where Portuguese had the highest rate of solving each problem (see Table \ref{tab:quant-completion-summary-2}) and also the highest rate of solving it in native language (see Table \ref{tab:quant-category-summary}).

\subsubsection{Writing Native English Code}

While the participants are not native English speakers, many of them described how they had a lot of experience writing code in English and that trying to think about writing code in their own language presented a unique challenge. For example, PC-6 highlights how variable names did not translate well from Chinese to English: 

\begin{quote}
    \textit{``I think the advantage is using my native language, I can describe the situation better and the disadvantage may be the variable name is English and it can't be translated accurately.''} (PC-6)
\end{quote}

Similarly, PC-9 claimed that \textit{``Certain words don’t exist''}, which suggests that even when writing in their native language, participants had to incorporate English terminology. PC-13 echoed this sentiment, explaining that naming conventions in code still had to conform to English standards, regardless of the language used for other parts of the input:

\begin{quote}
    \textit{``It behaves quite well in both English and Mandarin. but it have to make me explicitly name the function name so that it can pass.''} (PC-13)
\end{quote}

Several participants noted how deeply embedded the English language is in the context of programming, making it feel more natural to solve coding problems in English. This experience was captured by PC-3 as they said that trying to solve coding problems in their native language felt more like translation than direct problem-solving: 

\begin{quote}
    \textit{``I am really used to solve programming problems in english, answers, documentation, tutorials, youtube videos are mostly in English, trying to write them in my own language felt like I was translating my own thoughts. However writing prompts helps me have a look at the overall objective of a task and try to be precise which can be frustrating at first, but I guess it is useful in the long run.''} (PC-3) 
\end{quote}

PA-3 reiterated this point simply saying that programming is in English and therefore it is easier to write what you want in English.  

\begin{quote}
    \textit{``In English, it's easy to write what you want because the programming is in English, while in Arabic, it was a bit difficult to convey the information to ChatGPT''} (PA-3)
\end{quote}


%
%

\section{Discussion}
\label{sec:discussion}


In this work, we investigated students' success when using their non-English native languages to solve Prompt Problems (RQ1) and examined how using native-language prompts influenced their experience (RQ2). While it is now widely known that ChatGPT and other LLMs are highly effective at solving programming problems \cite{prather2023robots} -- particularly at the introductory level \cite{finnieansley2023my, finnieansley2022robots} -- their effectiveness with non-English prompts remains underexplored. As GenAI tools become increasingly integrated into computing education \cite{denny2024computingCACM}, understanding how they perform for non-English speakers is essential to fostering an inclusive learning environment.

Our findings indicate varying levels of success across language groups: students using Portuguese and Chinese achieved relatively high success rates, whereas Arabic speakers faced greater challenges in generating correct solutions.  In terms of perceptions, although many students felt more able to express themselves using their native language, they often thought that using English was better for solving the Prompt Problems citing better performance and closer alignment with programming constructs.  Several possible explanations may account for the challenges faced by students in this study. 

\subsection{Challenges of Non-English Prompting}

Firstly, although state-of-the-art LLMs have been shown to have multilingual capabilities, they are most capable in English due to the simple fact that the vast majority of the data they have been trained on is in English \cite{liu2024translationneedstudysolving}. In fact, although information on how proprietary models work internally is difficult to find, it appears that most generative AI models will first translate a non-English prompt into English before attempting to answer it \cite{stokel2024chatbot}.  As a result of this English-centric bias, it is not surprising that models may be less effective at determining the intent of non-English prompts \cite{dey2024betteraskenglishevaluation, zhao2024llamaenglishempiricalstudy}.  Our results align with prior research that shows LLMs perform better with high-resource languages where more training data is available.  Specifically, Arabic language data makes up less than 1\% of the language distribution available in Common Crawl\footnote{https://commoncrawl.github.io/cc-crawl-statistics/plots/languages} (an open repository of web crawl data that makes up a significant portion of the training data for many LLMs), whereas Chinese is one of the most common languages in the corpus (although still much less frequent than English).  This model bias may help to explain the particular difficulty faced by the Arabic speakers in our study.

Secondly, the output generated by the model is in a programming language that relies heavily on English syntax, which reduces the number of transformative steps needed when the input prompt is also in English. This alignment between the input (prompt) and output (program) language may also partially explain why students found writing prompts in English more effective.  Related to this point, programming languages inherently encode concepts naturally expressed in English, for example, constructs like conditional branching using keywords such as \texttt{if} and \texttt{while}.  These programming language keywords carry cultural and linguistic associations that may not translate directly into other languages \cite{becker2019parlezvous}.

Finally, the widespread use of English in programming has led students to become accustomed to writing code in English, even when it is not their native language.  While students reported that it was easier to describe their thoughts in their native languages, they still found themselves relying on English technical terms, making the shift back to their native language feel cumbersome and unnatural.  Thus, rather than simplifying the process, LLMs may be introducing an additional burden: students now have to translate their thoughts and intentions, as well as the syntax and technical terms. The first type of translation -- converting their thought process into code -- highlights how deeply embedded English is within programming, to the point where some participants are essentially thinking in a `coding language' closely tied to English.








\subsection{Implications for Practice}

Despite the challenges we observed and have hypothesized, our findings also highlight the potential for using the language translation capabilities of LLMs to make Prompt Problems -- and other kinds of new pedagogical assessments using GenAI -- more broadly and globally accessible.  This enables students to interact with programming concepts in their native languages, lowering barriers to entry and providing immediate access to programming activities and problem-solving tasks without being constrained by their English language abilities.

Our work contributes to the growing body of literature on broadening access to programming education for students with limited English proficiency.  Kumar's \textit{Refute} questions, for example, offer an alternative to Explain in Plain English (EiPE) questions by requiring students to identify incorrect logic rather than having to articulate explanations in English \cite{kumar2021refute, agarwal2023bugs}.  Recent work by Smith et al. demonstrates that Code Generation Based Grading can allow students to answer code comprehension questions in multiple languages, including with high correctness rates across several Indic languages~\cite{smith2024explainplainlanguagequestions}.  Similarly, we have observed that Prompt Problems offer students from diverse linguistic backgrounds a way to engage meaningfully with programming concepts in their preferred languages, promoting a more inclusive learning environment.  While our data suggests that students were often frustrated by how intertwined English is with programming itself, our data also show that the radical leap in the abilities of GenAI over the last few years could mean students who grow up with it do not feel the same tensions. It's possible that the student of the future does not think programmatically in English because of these advances.

There is growing momentum in the field toward making computing education more culturally relevant by contextualizing the problems that are solved in ways that are relevant to the local community \cite{coenraad2022using, Leonard_Sentance_2021, cota2022culturally}. 
However, the Prompt Problems used in this study were largely abstract and lacked deep cultural relevance
--- it would be interesting in future work to explore the effectiveness of Prompt Problems in English and non-English environments where the problem is more closely aligned to the non-English culture and where there may be more natural ways to describe the problem in the non-English language.
Designing problems that reflect culturally specific concepts and idioms may make non-English prompts more intuitive and accessible, enhancing students' learning experiences.

\subsection{Limitations}
\label{sec:limitations}

This study has several limitations that could be addressed in future work. First, the three groups of learners differed in their educational backgrounds: the students prompting in Chinese were postgraduates, while those prompting in Portuguese and Arabic were undergraduates. This variation may have influenced the results, as postgraduate students are typically more experienced with programming concepts. Second, the programming languages of instruction were not the same across the groups. Students prompting in Arabic generated code in Java, while those prompting in Chinese and Portuguese generated code in Python. While these differences in programming language could influence performance, we believe the findings remain valid, as current AI models are proficient at generating correct code in multiple popular languages, including Java and Python. Moreover, this reflects an interesting strength of the study, as the results suggest that non-English prompting can work across different programming languages.

Third, the study focused on three languages -- Arabic, Chinese, and Portuguese -- which limits the generalizability of our findings to other languages. Variations in the linguistic structure and the representation of languages in the training data of AI models could lead to different outcomes for students whose native languages were not included in this study. Further research is needed to explore the effectiveness of native-language prompting in a broader range of languages, including those classified as low-resource languages.

Finally, the sample sizes for each language group were relatively small, and additional factors, such as individual language fluency, prior programming experience, and familiarity with AI tools, could influence students’ success. These factors should be further investigated in future work.


%
%
\section{Conclusion}

English has long been the primary language of programming instruction, with the keywords and syntax of most popular programming languages being English-centric.  This has created a widely acknowledged barrier for non-native English speakers who must navigate both linguistic and technical challenges in computing courses.  With the rise of generative AI (GenAI), and in particular the language translation capabilities of modern large language models, there is potential to lower this barrier by allowing students to approach problem-solving in their native languages.  To investigate this potential, we introduced learners fluent in Chinese, Arabic, and Portuguese to Prompt Problems, a new kind of task in which programming exercises are solved by crafting natural language prompts instead of writing code directly. We found that students working in Portuguese and Chinese were generally able to achieve high success rates on these tasks, often expressing that native-language prompting allowed for a more natural articulation of their intent. In contrast, Arabic-speaking students appeared to face greater challenges, with lower accuracy and higher model misinterpretations.  This finding may reflect the limited quantity of training data in Arabic that modern GenAI models have been trained on.  Across all language groups, we observed a balance between expressivity and model performance -- students often found it easier to formulate ideas in their native language, but English often yielded more accurate responses, especially for programming terms embedded in the English-language syntax.  While we expect GenAI capabilities to continue to improve over time, especially for currently under-represented languages, our findings illustrate the transformative potential of existing multilingual GenAI tools. They can serve as valuable aids in democratizing programming education, making it more accessible and engaging for diverse learners globally.

\begin{acks}
This research was supported by the Research Council of Finland (Academy Research Fellow grant number 356114).
\end{acks}

\balance
\bibliographystyle{ACM-Reference-Format}
\bibliography{main}

\end{document}